**Exploring Strategies for Personalized Radiation Therapy: Part I – Unlocking Response-Related Tumor Subregions**

Hao Peng, Steve Jiang and Robert Timmerman

Department of Radiation Oncology, The University of Texas Southwestern Medical Center, Dallas, TX 75390, USA.

**Corresponding Author:**

Hao Peng, PhD, Email: Hao.Peng@UTSouthwestern.edu

**Running title:** Personalized Radiation Therapy

The authors have no conflicts to disclose.

Data sharing statement: The data that support the findings of this study are available from the corresponding authors upon reasonable request.

Author Contributions: Hao Peng: Methodology, Software, Formal Analysis, Investigation, Writing-Original Draft, Review & Editing. Steve Jiang and Robert Timmerman: Methodology, Review & Editing.

**Abstract**

Personalized precision radiation therapy requires more than simple classification—it demands the identification of prognostic, spatially informative features, and the ability to adapt treatment based on individual response. This study compares three approaches for predicting treatment response: standard radiomics, gradient-based features, and convolutional neural networks (CNNs) enhanced with Class Activation Mapping (CAM). We analyzed 69 brain metastases from 39 patients treated with Gamma Knife radiosurgery. An integrated autoencoder-classifier model was used to predict whether tumor volume would shrink by more than 20% at a 3-month follow-up, framed as a binary classification task. The results highlight the model's strength in hierarchical feature extraction and the classifier's discriminative capacity. Among the models, pixel-wise CAM provided the most detailed spatial insight, identifying lesion-specific regions rather than relying on fixed patterns—demonstrating strong generalization. In non-responding lesions, the activated regions may indicate areas of radioresistance. Pixel-wise CAM outperformed both radiomics and gradient-based methods in classification accuracy. Moreover, its fine-grained spatial features allow for alignment with cellular-level data, supporting biological validation and deeper understanding of heterogeneous treatment responses. Although further validation is necessary, these findings underscore the promise in guiding personalized and adaptive radiotherapy strategies for both photon and particle therapies.

# Introduction

At the University of Texas Southwestern Medical Center (UTSW), we are exploring an innovative approach called PULSAR (Personalized Ultra-fractionated Stereotactic Adaptive Radiotherapy) [1-5]. Differing from standard fractionated stereotactic radiation therapy (fSRT), PULSAR is designed to deliver high-dose radiation at intervals of two to four weeks (**Fig. 1A**). The benefits include more personalized treatment adjustment, enhanced normal tissue recovery, and potential synergy with concurrent immunotherapy.

Both PULSAR and fSRT require personalization, given the significant variability in individual patient responses to treatment. While some patients experience tumor reduction, others do not (**Fig. 1B**). Factors, such as tumor radioresistance and characteristics associated with the tumor microenvironment, complicate response. Currently, treatment plans are typically adjusted based solely on changes in gross tumor volume (GTV) observed between the first and second MRI scans. However, this assessment relies heavily on the physician's expertise and experience, which may not accurately align with treatment response [6-7]. By employing image features and radiomics analysis, we can transition the decision-making process from empirical judgments to a more data-driven approach, identifying how to achieve optimal therapeutic outcomes.

We previously investigated two approaches for predicting treatment response—radiomics and gradient analysis—with each offering distinct strengths and limitations [8, 9]. Radiomics extracts high-dimensional features using transformations like wavelets, capturing multi-scale textural and structural features. Notably, radiomics generates thousands of features and uses multiple feature selection steps to prevent overfitting. However, its interpretability is limited, relying on statistical correlations rather than causative insights [10-16]. On the other hand, gradient analysis is a simpler method, highlighting the characteristics of tumor borders. It is computationally efficient and interpretable but struggles with capturing complex spatial dependencies or subtle patterns detectable by neural networks [17-19].

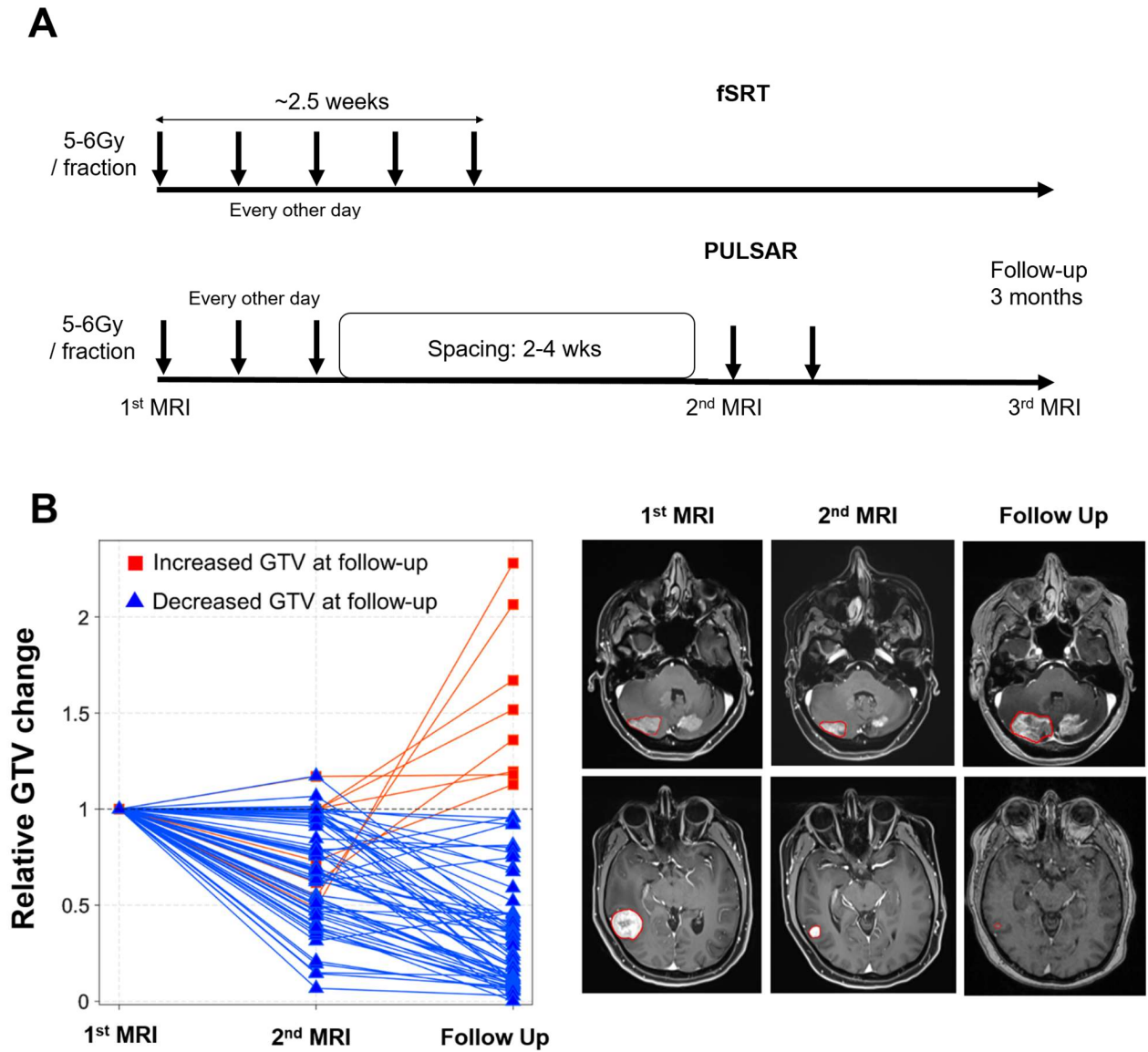

**Figure. 1. A.** Workflow comparison between fSRT and PULSAR. fSRT has the first and follow-up MRIs, while PULSAR extends the interval between two treatment cycles and adds a second MRI. **B.** Lesion volumes trajectories in PULSAR treatment at three time points (first MRI, second MRI, and follow-up). Two examples illustrate increased GTV after treatment (top) and reduced GTV (bottom) at follow-up.

In light of this, we were prompted to explore convolutional neural networks (CNNs), which are adept at capturing complex spatial patterns. Additionally, techniques like Class Activation Mapping (CAM) help improve interpretability by identifying image sub-regions critical for predictions [20-25]. Given the complementary strengths of radiomics, gradient analysis, and CNNs, an intriguing question emerges: whether and how the features identified by these different methods correlate with each other? These correlated features could be leveraged to enhance predictive robustness, while uncorrelated features might offer complementary insights, contributing to a multi-dimensional predictive model.

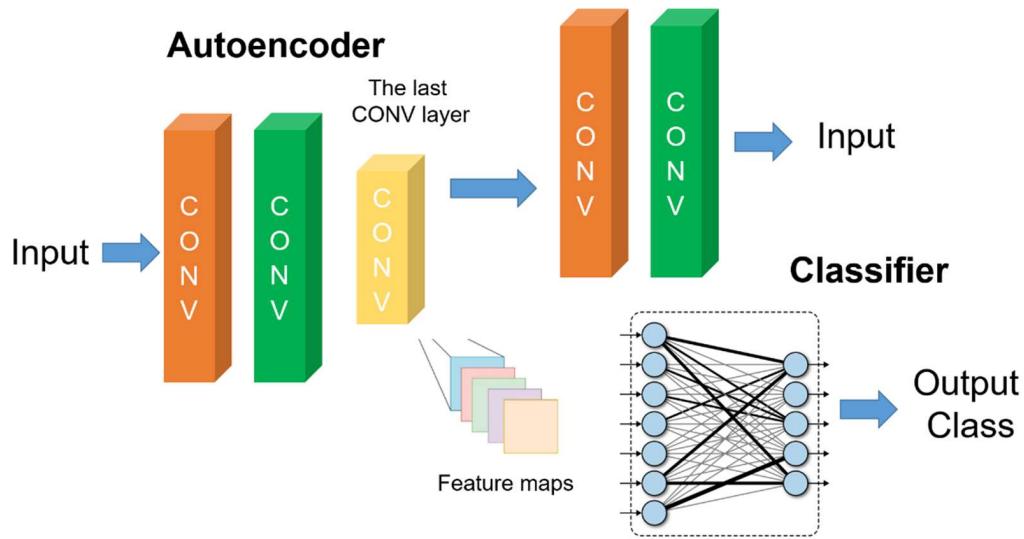

**Figure. 2.** The combination of an autoencoder and a classifier to perform binary classification and derive CAMs (Class Activation Maps).

**Results**

**Data processing and CNN modeling structure.** We analyzed preliminary results from a cohort of brain metastasis (BMs) treated with PULSAR. Details of patient data collection are provided in **Table S1**. The prediction task was framed as a classification problem to identify lesions in two groups: non-response (Group A) and response (Group B), based on whether lesions exhibited a $\geq 20\%$ volume reduction in 3-month follow-up MRIs [26-28]. Our previous publication provides details on analyses conducted using standard radiomics and gradient features [8, 9]. As shown in **Figure 2**, the integration of an auto-encoder and classifier was used for image reconstruction and classification simultaneously. The autoencoder comprised convolutional and pooling layers for feature extraction and a bottleneck layer for dimensionality reduction. The attached classifier consisted of dense layers with batch normalization and dropout, outputting binary predictions via sigmoid activation. The model was trained on 64×64×1 image inputs (2D) through simultaneous fine-tuning of reconstruction and classification tasks. Although sequential training was explored, its performance was found to be inferior. Accuracy and loss were monitored on training and validation sets, with periodic visualization of reconstructed images and feature maps to track progress. Performance metrics, including sensitivity, specificity, accuracy, precision, F1 score, and AUC, were compared with the radiomics and gradient-features methods in our previous studies [8, 9].

After training, three CAM variants were applied to highlight regions contributing to classification decisions: Grad-CAM, Grad-CAM++, and pixel-wise CAM (see details in the Supplemental Materials). The primary differences among the three variants lie in how they compute regional importance and localize relevant features. The last convolutional layer of the autoencoder, comprising five feature maps (16×16×5), captures high-level semantic information relevant to the activated class while retaining spatial resolution. CAM aggregates learned patterns from feature maps to generate heatmaps that highlight meaningful, high-level semantic content, anchored in distinguishing between response and non-response classes.

**Feature extraction for standard radiomics and gradient-based radiomics. Figure 3** illustrates examples of feature extraction for standard radiomics and gradient-based methods. Through wavelet transforms, such as high- and low-frequency filtering, radiomic features capture the global texture details across the entire gross tumor volume (GTV) (**Fig. 3A**). Two top features identified in our studies are low gray-level zone emphasis (LGLZE) which quantifies the prevalence of low gray-level size zones, and kurtosis which probes tumor heterogeneity [8]. Both features provide just a single value for the entire GTV without explicit spatial information. In gradient-based features (**Fig. 3B**), analysis is performed for both core and margin to incorporate more spatial information [9]. For example, gradient magnitude quantifies the rate of intensity change at each voxel, with high values indicating significant changes. Radial gradient measures the alignment of the gradient vector with the radial direction from the center of mass to the voxel, helping to understand the directional characteristics of tumor growth or shrinkage. Gradient-based features pinpoint *where* abrupt changes occur—providing location-sensitive information that is often linked to treatment response or progression patterns. For example, in the context of brain metastases, a steep gradient at the tumor boundary may indicate a well-defined lesion, which is often associated with a better prognosis. In contrast, a diffuse or low-gradient margin could signal infiltrative growth or peritumoral edema, which may correlate with resistance to therapy or higher recurrence risk.

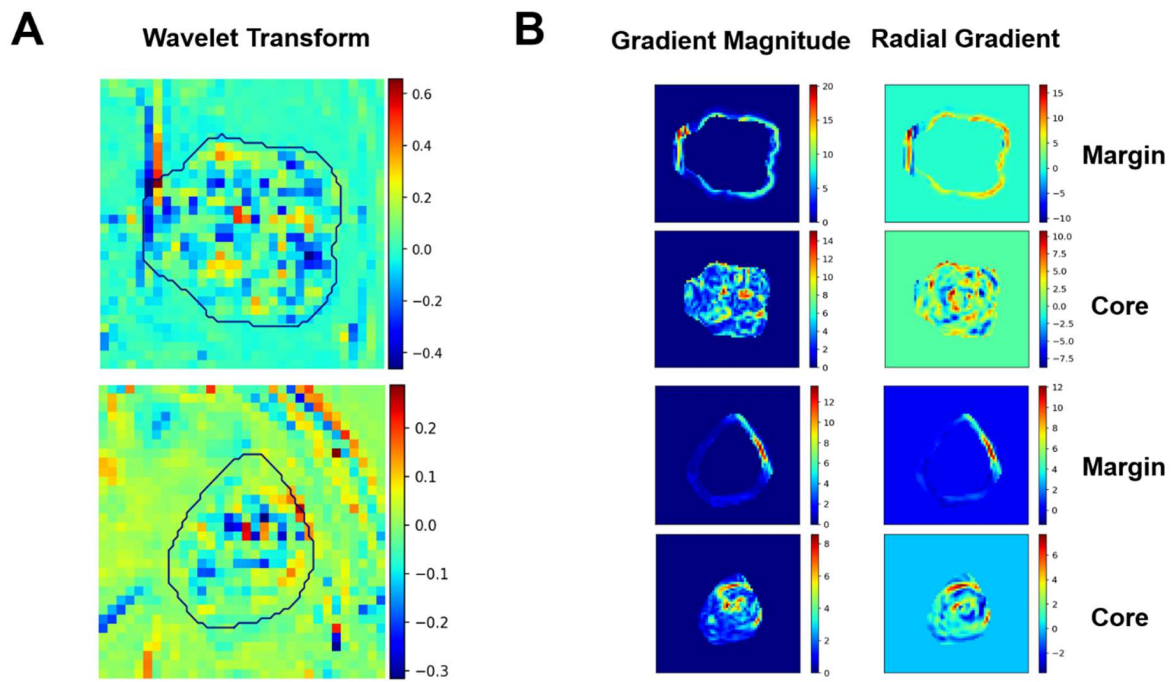

**Figure. 3.** Illustration of how to extract conventional radiomics and gradient features for two lesions. **A.** For conventional radiomics, wavelet transform is applied for the derivation of the gray-level co-occurrence matrix (GLCM) and texture metrics. **B.** Gradient features, including gradient magnitude (GM) and radial gradient (RG), are calculated for the core and margin of the lesions.

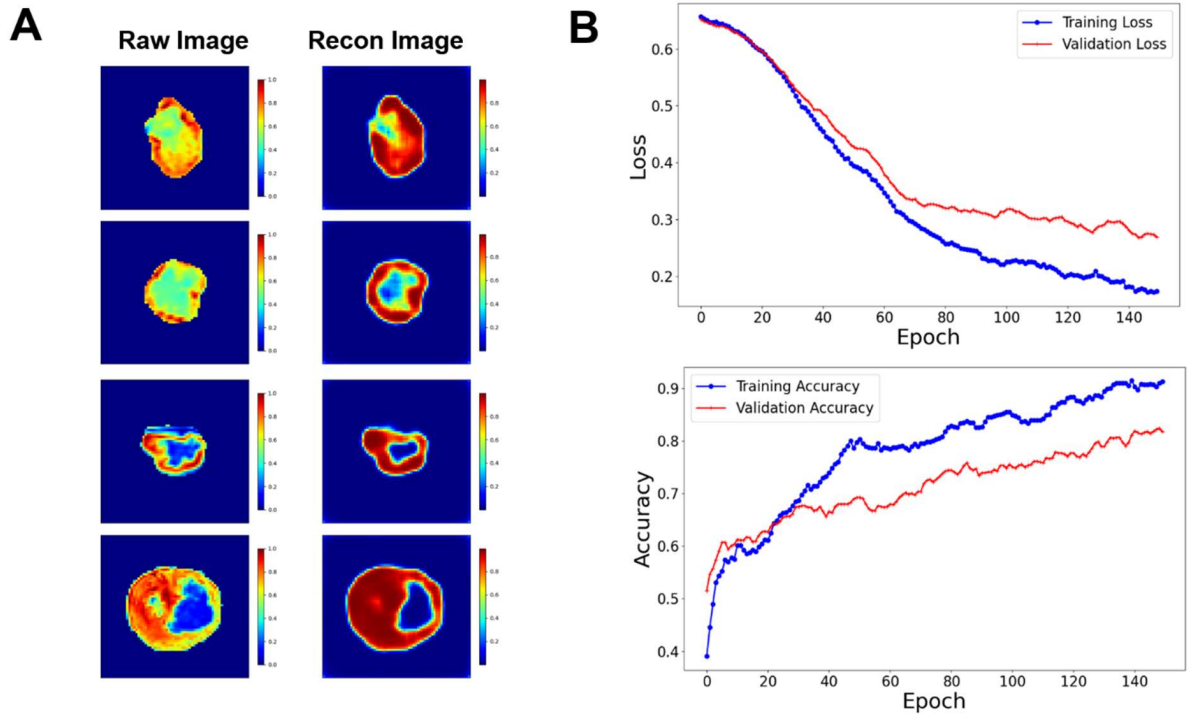

**Figure. 4. A.** The raw image alongside the reconstructed images (output of the auto-encoder). **B.** The loss curve and accuracy curve, plotted as functions of the number of epochs during training.

**CNN-based classification and CAM analyses.** For CNN-based classification, **Fig. 4A** illustrates the raw image and reconstruction for two selected lesions, demonstrating that the autoencoder successfully extracts hierarchical features. **Fig. 4B** shows that the accuracy and loss curves, trained on the first MRI set and tested on the second MRI set, converge around epoch 150 and show no signs of overfitting.

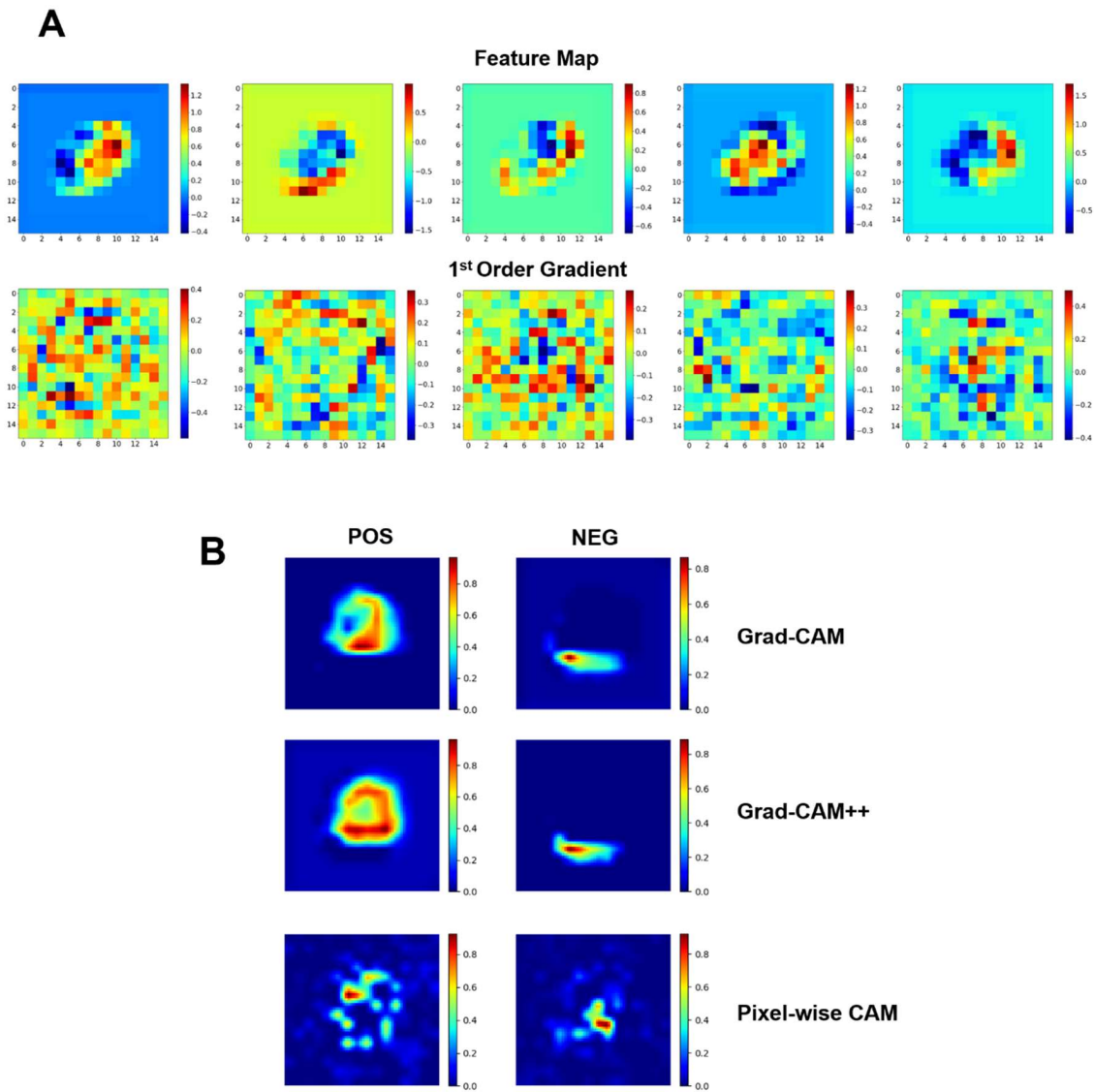

**Figure. 5. A.** Five feature maps and the first-order gradient maps. **B.** Examples of three CAM variants. Each column shows the contribution of each pixel for the classifier's decision to the true class (POS) and the incorrect class (NEG), respectively.

**Fig. 5A** shows the five feature maps and the first-order gradient (i.e., class score relative to the features in the feature maps). **Fig. 5B** presents heatmaps of Grad-CAM, Grad-CAM++, and pixel-wise CAM for a selected lesion. CAMs highlight specific regions that the model focuses on to classify the input image, revealing areas that strongly contribute to the model's decision on whether a lesion belongs to the true class (***POS***), or the wrong class (***NEG***). In both Grad-CAM and Grad-CAM++, averaging the gradients

Each feature map generates a smoother activation map that highlights bulk regions, either within the tumor or along its boundary—resembling the gradient-based patterns observed in **Fig. 3B.** In contrast, the pixel-wise CAM heatmap shows more fine-grained details and multiple small niches. Since tumor structures differ significantly from object classification in natural images due to factors such as intra- and inter-tumoral heterogeneity, pixel-wise CAM was chosen for the subsequent analyses.

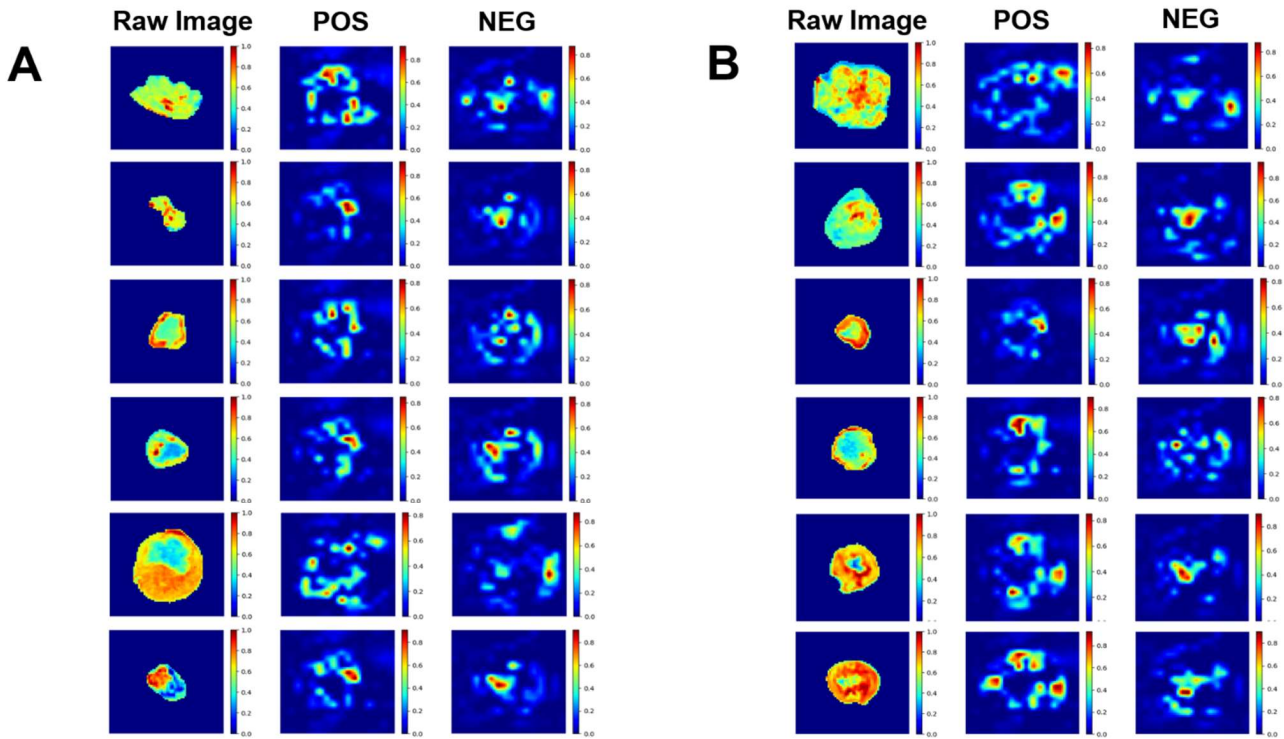

**Figure. 6.** The pixel-wise CAMs for six lesions from each group, non-response (**A**) and response (**B**).

**Figure 6** presents the pixel-wise CAMs for six lesions from each group: non-response (Group A) and response (Group B). The variation in highlighted regions suggests that the model has learned distinct image patterns that effectively differentiate between the two groups. Notably, the activation maps avoid repeatedly focusing on fixed areas, instead emphasizing lesion-specific regions—demonstrating strong generalization. For non-responding lesions, the activated areas may correspond to sub-regions of radioresistance, potentially indicating the need for dose escalation. Additionally, the maps do not exclusively target either the lesion core or boundary, implying that features from both regions contribute to classification. **Table 1** presents a comparison of classification performance across the three methods, showing that the CAM-based model outperforms the others. To account for class imbalance, we report the F1 score alongside other evaluation metrics for a more balanced performance assessment [29, 30]. It should be noted that in our previous studies, both radiomics and gradient-based features derived from the second MRI (i.e., delta features) helped improve prediction accuracy. However, in this work, we limited the information to the first MRI only to prioritize early decision-making at the onset of treatment.

**Table 1. The performance metrics of different models for the test dataset (2$^{nd}$ MRI).** The mean and standard deviation are calculated from 50 trials using multiple 80%/20% training/test set splits, consistent with our previous studies [8, 9].

|  | Radiomics Model | Gradients Model | Pixel-CAM Model |
|---|---|---|---|
| **Sensitivity** | 0.503 ± 0.271 | 0.779 ± 0.267 | 0.877 ± 0.045 |
| **Specificity** | 0.745 ± 0.128 | 0.723 ± 0.133 | 0.738 ± 0.39 |
| **Accuracy** | 0.696 ± 0.107 | 0.733 ± 0.112 | 0.807 ± 0.027 |
| **AUC** | 0.748 ± 0.143 | 0.826 ± 0.111 | 0.844 ± 0.03 |
| **Precision** | 0.344 ± 0.197 | 0.434 ± 0.186 | 0.771 ± 0.027 |
| **F1 Score** | 0.389 ± 0.193 | 0.541 ± 0.187 | 0.820 ± 0.027 |

**Discussion**

Personalized precision radiation therapy goes beyond classification. CAM-based approach focuses on identifying the most prognostic features and maintaining spatial information, potentially allowing for timely adjustments to treatment plans, such as dose escalation or surgery. It strikes a balance in the trade-off between localization and interpretability. The feature maps from the last convolutional layer capture higher-level characteristics (e.g., edges, textures, or object parts) within the GTV. CAM models use these feature maps to study which sub-regions of the input image most influenced the model's prediction. We speculate that the presence of multiple distinct regions within a tumor may indicate radio-resistance, due to variations in structural heterogeneity, oxygen levels, blood supply, biological traits, and genetic makeup. Such speculation could be tested with further investigation into whether these regions correspond to tumor biology or treatment-specific patterns. For example, these regions could be compared with pathological or molecular patterns, such as those identified through single-cell resolution CODEX (CO-Detection by indexing) [31-33] or pixel-level spectroscopic MRI (MRS) [34]. CODEX can provide detailed information about the molecular and cellular makeup of the tumor, such as the expression of specific genes, proteins, and immune markers. This information can help identify tumor heterogeneity and the presence of resistant subpopulations within different regions of the tumor. MRS can help measure metabolic activity (e.g., lactate, choline, lipids) and oxygenation levels within the tumor. Regions with low oxygen levels (hypoxia) or high metabolic activity are often more resistant to radiation therapy.

Leveraging identified sub-regions, rather than relying solely on global classification outcomes, may help link imaging features with pathomics for personalized treatment and enhanced biological insight. Recently, both registration-based and non-registration approaches have shown promise in integrating macroscopic image features with microscopic biology. Although MRI and pathology differ vastly in spatial resolution—approximately 0.5 mm for MRI versus 20 microns for pathology—methods correlating them from either broadly corresponding or perfectly co-localized regions have emerged as effective tools for virtual biopsies. For example, Brancato et al. analyzed MRI-derived radiomic features, such as Apparent Diffusion Coefficient (ADC) and T1 contrast, alongside pathomic characteristics in glioblastoma patients [35]. By applying radiomics on cell density maps, they found meaningful correlations—for instance, an inverse relationship between mean ADC and nuclei count. Bobholz et al. employed a more spatially precise approach by co-registering postmortem multi-sequence MRIs directly with histological images in brain cancer patients [36]. Their model predicted histology-derived cellularity from localized MRI intensities, successfully mapping hypercellular regions confirmed by immunohistochemical markers of proliferation and microvascular density. This precise spatial registration enabled direct association of imaging features with specific histological sub-regions, deepening biological interpretability. Building on these advances, identifying tumor sub-regions using the CAM techniques allows us to investigate whether imaging features extracted over these localized areas correlate with pathomic patterns, uncovering the biological mechanisms driving treatment response. This would be a critical step toward building "virtual biology" models that can noninvasively infer tissue characteristics from imaging data.

Understanding the differing performance of three CAM models requires further investigation and validation. Grad-CAM computes a single global importance weight for each feature map (channel), effectively answering the question: On average, how much does this entire feature map contribute to the target class? However, this averaging can overlook fine-grained or small localized features. Grad-CAM++ improves Grad-CAM by computing a separate weight for every pixel within each feature map, utilizing first-, second-, and third-order gradients. This allows the model to identify which specific parts of the feature map contributed to the prediction, rather than treating the entire map uniformly. The pixel-wise weights are then aggregated into a single scalar per feature map, like Grad-CAM, but with spatial sensitivity. As a result, Grad-CAM++ can provide better localization of small regions contributing to the same class or multiple instances of objects from the same class. Nonetheless, higher-order gradient terms can sometimes be small or near zero due to neural network properties. This might explain why comparable results are observed between Grad-CAM and Grad-CAM++ in **Fig. 5B**. In contrast, pixel-wise CAM skips averaging or global weighting altogether and directly produces heatmaps by using pixel-wise gradients. This preserves full spatial detail and results in sharper, more detailed heatmaps. However, it tends to produce noisy or fragmented maps and is sensitive to small input perturbations, which can reduce its robustness.

The optimal CAM method for treatment outcome prediction in our task remains to be determined. In natural images, where features are often abundant, complex, and spatially dispersed—such as overlapping objects and varying textures—Grad-CAM++ shows robust performance [24]. In contrast, tumor images typically exhibit more defined and consistent morphology, with clear boundaries and recognizable structural patterns. These spatial features are less cluttered, well-organized, and less noisy. When the analysis is restricted to the GTV by masking irrelevant regions as done in our study, we anticipate that pixel-wise CAM will yield more precise and meaningful heatmaps. Furthermore, the regions identified by CAM can be compared with simple gradient-based radiomic features—such as those capturing spatial heterogeneity and edge contrast—to establish correlations (**Fig. 3**). Correlating CNN-derived attention maps with handcrafted gradient-based features will provide a useful cross-validation approach.

Due to the small cohort size, efforts were made to minimize overfitting in our study. This included incorporating regularization techniques such as dropouts in the CNN model. Dropout works by randomly "dropping out" a fraction of neurons during each training iteration, temporarily removing them from the network so they do not contribute to the output. For training, the first MRI scan was used, while the second MRI was reserved for performance evaluation assuming that meaningful spatial features remain stable over time. However, by the time of the second treatment cycle (~3 weeks later), the tumor has undergone changes. That said, the validation is likely to be more stringent than using samples from the first MRI. This also explains the comparatively lower performance observed on the validation dataset, as shown in **Fig. 4B**. With more data and evaluation performed using the first MRI, the classification performance reported in **Table 1** is expected to improve slightly. Additionally, compared to standard radiomics and gradient-based radiomics, it is important to note that using either delta values (i.e., change of features between pre- and post-treatment) or ensemble models incorporating dosiomics features, can dramatically enhance performance and achieve sensitivity and specificity over 90% [8, 9]. However, these approaches still lack the ability to localize the spatial regions responsible for the observed treatment responses. Earlier decision-making at the onset of treatment is only possible by leveraging the first MRI.

Several challenges remain that should be addressed moving forward. First, the dataset used in this study was small and required data augmentation during training. Further validation on larger datasets is necessary to confirm our preliminary findings. Second, deep features extracted through convolutional layers are often considered a "black box," as the abstract patterns learned by the network do not always correspond directly to clinically or biologically interpretable concepts. Assigning meaningful clinical or biological significance to these features remains a major challenge. Without additional validation, it is hard to determine whether the extracted features represent genuine biological signals or spurious correlations. Differentiating true tumor progression from pseudo progression or radiation necrosis is difficult because these conditions exhibit overlapping imaging characteristics. This overlap complicates volumetric assessments and weakens the link between spatial features and classification labels. Third, this study focused solely on predicting GTV change at three months post-treatment as a classification task. Future research should investigate how these volume changes correlate with local control and other long-term outcomes. Additionally, the class separation and CAM results may be affected by the choice of a 20% tumor volume change as the response threshold. Exploring alternative thresholds (e.g., 30% or 50%) could offer deeper insight into the robustness and generalizability of the findings. Finally, to assess the framework's broader applicability and robustness, it should be tested on other tumor types and locations. CAM results may be more complex for glioblastoma multiforme (GBM) compared to brain metastases, due to its infiltrative growth patterns and less defined tumor margins.

**Conclusion**

Radiation therapy should not be a one-size-fits-all approach. Our study shows the potential of CNN and CAM models in predicting treatment response and identifying tumor sub-regions potentially associated with radioresistance. It demonstrated superior predictive capability compared to conventional radiomics and gradient-based methods, striking an effective balance in the trade-off between localization and interpretability. Further investigation into the potential correlation between our method and spatial biology may provide additional insights. This approach holds the potential to transform personalized radiation therapy by integrating diagnostic imaging with complementary pathological and molecular insights, paving the way for more precise and biologically informed treatment strategies.

## Methods

**PULSAR data collection.** PULSAR patients with BMs were treated utilizing Gamma Knife Icon™ (Elekta AB, Stockholm, Sweden). Patients first undergo a pretreatment MRI scan followed by the initial treatment course, which consists of three fractions or pulses (5 to 6 Gy per fraction/pulse) with a two-day interval between fractions. Subsequently, after a span of about three weeks, the second treatment cycle is administered based on the second MRI scan, allowing for adjustments to changes. Our retrospective study focused on 39 patients who underwent PULSAR treatment at UTSW. This cohort included 69 lesions treated between November 1, 2021, and May 1, 2023, encompassing patients with both single and multiple metastases. Detailed demographic and clinical profiles, such as age, gender, histology, lesion number, and treatment specifics, are provided in **Table S1**. During the treatment process, the gross target volume, clinical target volume, and planning target volume were considered equivalent, as no margin expansions were applied. Lesions with a Gradient Index (GI) greater than 4 were excluded from the study. We confirm that the study was approved by our Institutional Review Board (IRB), and all methods were performed in accordance with the relevant guidelines and regulations. This study is retrospective in nature, and the requirement for informed consent was waived.

We collected MRI images and radiotherapy contour structure files (RTstructure). The MRI images were acquired using axial (AX) sequences with T1-weighted enhancement. Tumor volumes in follow-up MRI images (without enhancement) were assessed three months after PULSAR treatment. Two board-certified radiation oncologists conducted a thorough comparison of MRI images and GTV contouring for each lesion. Our goal is to predict whether a lesion will demonstrate a volume reduction of ≥ 20% at follow-up after undergoing PULSAR treatment. Previous studies have demonstrated a strong correlation between a volume reduction of 20% or more and improvement in neurological signs and symptoms [26-28]. We applied the same criterion and framed it as a classification problem. Tumors with follow-up volumes at or above 80% of their initial volume were categorized as "non-response" (referred to as Group A), while those with volumes reduced below this threshold were categorized as "response" (referred to as Group B).

**AI model development and training.** We built an autoencoder-based pipeline capable of accurate image reconstruction, with an attached classifier for prediction, using the Keras package (https://github.com/keras-team/keras). The autoencoder consists of an encoder-decoder structure, designed with convolutional and pooling layers. The bottleneck of the encoder generates feature representations of reduced spatial resolution, which aids in capturing essential features while discarding unnecessary details. The size of image input is 64×64×1. The encoder consists of two convolutional layers with ReLU activation and 3×3 kernels, each followed by a 2×2 max-pooling layer for spatial downsampling. These layers capture hierarchical features, reducing the input dimension down to a 16×16×5 representation of the bottleneck. The bottleneck layer applies a mask, ensuring that features learned from outside of the GTV in the input are suppressed. This masked feature map is then passed to the decoder. The decoder uses two upsampling layers with corresponding convolutional layers to restore spatial resolution to the input size. The final layer outputs a single-channel grayscale image using a sigmoid activation to produce pixel intensity values between 0 and 1. The loss function is designed to balance perceptual fidelity using Structural Similarity Index (SSIM) (weight: 0.2), and pixel-wise accuracy using binary cross-entropy (BCE) (weight: 0.6), plus a L2 regularization term of the feature maps (weight: 0.2). SSIM measures the perceptual similarity between predicted and ground truth images, encouraging the autoencoder to maintain structural information. BCE provides pixel-wise error minimization, capturing the pixel intensity differences. The penalty term is added that discourages non-zero activations in masked (zero-valued) regions of the encoded features. This additional penalty helps the model generalize better by focusing only on a small number of features.

The classifier consists of one dense layer (N=32) with batch normalization, and dropout regularization, followed by a fully connected layer with sigmoid activation to output binary predictions. The batch normalization layer stabilizes training, while the dropout layer (ration=0.1) helps mitigate overfitting. It is worth mentioning that Global Average Pooling (GAP) or Global Max Pooling (GMP), is not used here because the dataset size is small so that the model cannot afford more feature maps (e.g., channel-wise information). Instead, we have a dense layer before the fully connected layer, which serves the purpose of reducing feature dimensionality like principal component analysis (PCA). The dense layer learns to combine and weigh these inputs in various ways, capturing interactions between features derived from different feature maps. The final fully connected layer outputs class probabilities based on the learned feature representations.

We investigated training these two tasks simultaneously (batch size: 64), with the losses from the autoencoder and classifier weighted equally. As an alternative, we also tested a sequential approach: first training the autoencoder, freezing the encoder weights, and then training the classifier. However, this approach demonstrated inferior performance and is therefore not included in the manuscript. Throughout training, accuracy is monitored on both the training and validation sets. Progress is periodically visualized by reconstructing sample images and plotting feature maps. Additionally, classification accuracy and loss are tracked for both training and validation data, providing insights into the model's performance and potential overfitting. During training, dropout was used to randomly "drop" certain neurons, forcing the network to generalize. Data augmentation was introduced by rotating the raw image at different angles

(10°, 15°, 30°, 45°, and 90°), yielding a total of 414 samples (69×6=414). The second MRI dataset (69 samples without data augmentation) was used for validation and performance evaluation, with dropout disabled. The AdamW algorithm with default settings was used for optimization (learning rate = 0.0005, weight decay = 0.0001, beta1 = 0.9, beta2 = 0.999, epsilon = $1\times10^{-8}$).

**CAM visualization.** CAMs are class-discriminative saliency maps, providing information on the most important parts of an image for identifying a particular class. Mathematically, the class score for an output class $c$ (either 1 or 0 in our study) in a standard CAM model with a Global Average Pooling (GAP) layer is [21-23]:

$$y^c = \sum_k w_k^c \frac{1}{Z} \sum_i \sum_j A_{ij}^k \qquad (1)$$

where $A_{ij}^k$ is pixel at location (i, j) in each feature map ($k$=1, 2, 3, 4 and 5 in our study), Z is the total number of pixels in the feature map, and $w_k^c$ is the weight of the $k^{th}$ feature map for class c. The final CAM can be represented by Eq. (2).

$$CAM^c = ReLU(\sum_k w_k^c A^k) \qquad (2)$$

Applying the ReLU function ensures that only the positive contribution to the class prediction is retained. The CAM can be shown as a heatmap in which the regions with more positive values are the most salient for a particular prediction, while the regions with more negative values give those patches in the image that adversarially affect a particular prediction. In our study, three versions were tested to obtain $w_k^c$:

- **Grad-CAM**: One limitation of a standard CAM (with the use of GAP) is the overhead of learning the weights [23]. Gradient-weighted class activation map (Grad-CAM) is a modification to overcome this limitation, by linking the weights directly to the derivative of the output class score with respect to the pixels $A_{ij}^k$ in the feature map. As shown in Eq. (3), Grad-CAM uses the global average of the gradients as weights, scaling all pixel gradients by the same factor 1/Z.

$$w_k^c = \frac{1}{Z}\Sigma_i \Sigma_j \frac{\partial y^c}{\partial A_{ij}^k} \qquad (3)$$

- **Grad-CAM++**: In Grad-CAM, averaging the gradient across spatial positions sacrifices spatial details because it condenses all the information from a feature map into a single importance weight [24]. To address this limitation, Grad-CAM++ aims to produce more detailed and accurate heatmaps, especially in scenarios involving multiple objects of the same class, effect of different footprints, or fine-grained interpretability. As shown in Eq. (4), Grad-CAM++ uses higher order gradients (second and third-order gradients).

$$w_k^c = \Sigma_i \Sigma_j \alpha_{i,j}^{k,c} ReLU\left(\frac{\partial y^c}{\partial A_{ij}^k}\right) \qquad \alpha_{i,j}^{k,c} = \frac{\frac{\partial^2 y^c}{(\partial A_{ij}^k)^2}}{2\frac{\partial^2 y^c}{(\partial A_{ij}^k)^2} + \Sigma_a \Sigma_b A_{a,b}^k \frac{\partial^3 y^c}{(\partial A_{ij}^k)^3}} \qquad (4)$$

- **Pixel-wise CAM:** Instead of averaging within pixels in a single feature layer to get a single value (e.g., $w_k^c$), the gradient for each spatial location is multiplied by the corresponding activation in the feature map. This results in a per-pixel weight that directly reflects the contribution of each spatial location to the output. Each pixel in the feature map is treated independently, leading to a more fine-grained localization of the important regions within the feature maps as shown in Eq. (5).

$$w_{i,j}^{k,c} = \frac{\partial y^c}{\partial A_{ij}^k} \qquad CAM^c = ReLU(\Sigma_k w_{i,j}^{k,c} A_{i,j}^k) \qquad (5)$$